\documentclass[12pt,preprint]{aastex}

\usepackage{natbib}
\usepackage{amsmath}
\usepackage{float}
\usepackage{subfig}

\begin{document}

\title{The Galactic Center Weather Forecast}
\author{M. Mo{\'s}cibrodzka$^{1\dagger}$, H. Shiokawa$^2$,
  C. F. Gammie$^{2,3}$, J. C. Dolence$^{4}$}

\affil{$^1$ Department of Physics and Astronomy, University of Nevada, 4505
  South Maryland Parkway, Las Vegas, NV 89154}

\affil{$^2$ Astronomy Department, University of Illinois, 1002 West Green
  Street, Urbana, IL 61801}

\affil{$^3$ Department of Physics, University of Illinois, 1110 West Green
  Street, Urbana, IL 61801}

\affil{$^4$ Department of Astrophysical Sciences, Princeton University, Peyton
Hall, 4 Ivy Lane, Princeton, NJ 08544}
\email{$^{\dagger}$ monikam@physics.unlv.edu}

\begin{abstract}

In accretion-based models for Sgr~A* the X-ray, infrared, and millimeter
emission arise in a hot, geometrically thick accretion flow close to the
black hole. The spectrum and size of the source depend on the black hole
mass accretion rate $\dot{M}$.  Since Gillessen et al. have recently
discovered a cloud moving toward Sgr~A* that will arrive in summer 2013,
$\dot{M}$ may increase from its present value $\dot{M}_0$.  We therefore
reconsider the ``best-bet'' accretion model of Moscibrodzka~et~al.,
which is based on a general relativistic MHD flow model and fully
relativistic radiative transfer, for a range of $\dot{M}$.  We find that
for modest increases in $\dot{M}$ the characteristic ring of emission
due to the photon orbit becomes brighter, more extended, and easier to
detect by the planned Event Horizon Telescope submm VLBI experiment.  If
$\dot{M} \gtrsim 8 \dot{M}_0$ this ``silhouette'' of the black hole will
be hidden beneath the synchrotron photosphere at $230$GHz, and for
$\dot{M} \gtrsim 16 \dot{M}_0$ the silhouette is hidden at $345$GHz.  We
also find that for $\dot{M} > 2 \dot{M}_0$ the near-horizon accretion
flow becomes a persistent X-ray and mid-infrared source, and in the
near-infrared Sgr~A* will acquire a persistent component that is
brighter than currently observed flares.

\end{abstract}

\section{Introduction}

The recent discovery of a cloud moving towards the Galactic Center
\citep{gillessen:2012} creates a potential opportunity for testing
models of Sgr~A*.  The $> 3 M_{\oplus}$ cloud will interact strongly
with gas near nominal pericenter at $r_p \simeq 300 {\rm AU} \simeq 8000
GM/c^2$ ($M \equiv$ black hole mass), and may change the black hole
accretion rate $\dot{M}$.  Since the structure of the cloud and the
surrounding medium are uncertain, possible outcomes range from very
small changes in the accretion rate over timescales of decades to rapid,
large changes in the accretion rate.

The dynamical timescale at $r_p$ is $t_d = (r_p^3/(G M))^{1/2} = 0.5$yr,
and the viscous timescale $t_{vis} = (\alpha \Omega)^{-1} (R/H)^2
\approx 100$yr $\gg t_{PhD}$, assuming $\alpha = 0.05$ and $H/R = 0.3$,
i.e. a hot, radiatively inefficient accretion flow.  After an initial
transient phase while the flow circularizes---accompanied by transient
emission---it is natural to think the flow will settle into a steady
state.  The settling timescale could be as little as a few $t_d$, and so
the steady state may arrive as soon as mid-2014.  If the resulting flow
can be modeled as a steady disk, the excess mass will drain away on the
viscous timescale, i.e. the source will remain bright well into the 22nd
century.  It is therefore interesting to ask how changes in $\dot{M}$
will manifest themselves observationally.

Current observations of Sgr~A* show $F_\nu = 0.5-1$ Jy at 1-50 GHz with
a nearly flat spectral slope \citep{falcke:1998}; $F_\nu \sim
\nu^{p=0.17-0.3}$. The spectral slope becomes flatter and variable at
230-690 GHz, with $p= -0.46 - 0.08$, \citealt{marrone:2006b}), which is
commonly interpreted as signaling a transition from optically thick to
optically thin synchrotron emission.  The discovery of polarized
emission (polarization fraction at level of a few per cent) at
$\lambda=1.3$mm and subsequent measurement of Faraday rotation imply a
model dependent limit $2 \times 10^{-7} > \dot{M} > 2 \times 10^{-9}
{\rm M_\odot yr^{-1}}$ in the inner accretion flow
(\citealt{bower:2005}, \citealt{marrone:2006a}).  Sgr~A* is resolved by
VLBI at $1.3$mm where interstellar scattering is comparable to the
intrinsic source FWHM $37^{+16}_{-10} \mu{\rm as}$
\citep{doeleman:2008}.  This is smaller than the apparent diameter of
the event horizon $\sim 55 \mu as$.  Sgr~A* fluctuates rapidly in the
near-infrared (NIR), with average $F_\nu \sim 1 {\rm mJy}$
(\citealt{dodds-eden:2011}).  It is not yet detected at mid-infrared
(MIR) wavelengths ($\lambda \leq 8.6 \mu {\rm m}$, e.g.
\citealt{schoedel:2011}).  In the X-ray Sgr~A* exhibits flares with
characteristic duration of order an hour and a duty cycle of order
$5\%$.  The upper limit for quiescent state X-ray emission is $\nu L_\nu
< 2.4 \times 10^{33} {\rm erg s^{-1}}$ \citep{baganoff:2003}.  We are
not aware of any secular trends in these observed properties of Sgr~A*.

One model of Sgr~A* that fits most observational constraints is our
relativistic accretion model \citep{moscibrodzka:2009}, where
submillimeter, IR, and X-ray emission arise in an optically thin,
geometrically thick disk close to the event horizon; radio emission is
assumed to arise nonthermally in a synchrotron photosphere at larger
radius \citep{ozel:2000}, but is not predicted by our model, which
focuses on the inner parts of the accretion flow.  The underlying flow
model is a general relativistic magnetohydrodynamic (GRMHD) simulation
(\citealt{gammie:2003}, \citealt{noble:2006}, \citealt{shiokawa:2011}).
The emerging radiation is calculated using Monte Carlo
\citep{dolence:2009} and ray-tracing schemes.  The $1.3$mm flux
originates as Doppler boosted synchrotron emission from the approaching
side of the disk between $\sim 10 GM/c^2$ and the innermost stable
circular orbit (ISCO).

Our relativistic disk model assumes a thermal electron distribution
function, neglects thermal conduction, and assumes a constant ratio of
ion to electron temperature $T_i/T_e$.  The model does not produce the
observed IR flaring at the correct amplitude, but can with the addition
of a small nonthermal component in the distribution function
\citep{leung:2010}. If we fix the black hole mass ($M_{BH}=4.5 \times
10^6 M_{\odot}$, \citealt{ghez:2008}) and distance ($D=8.4\, {\rm kpc}$,
\citealt{gillessen:2008}), the remaining model parameters are the source
inclination $i$, black hole spin $a_*$ ($0 \le a_* \le 1$), $T_i/T_e$,
and $\dot{M}$. We fix $\dot{M}$ so that the $1.3$mm flux matches the
observed $\simeq 3$ Jy.

The relativistic accretion models are not tightly constrained by the
data, but they reveal the following: (1) face-on models that reproduce
the millimeter flux would look like rings and therefore, in VLBI data,
have dips in visibilities on fixed intermediate baselines, while
existing observations suggest that the ring radius would need to vary to
fit the data (\citealt{fish:2011}; see also the discussion of
\citealt{broderick:2011}).  More nearly edge-on models are therefore
favored; (2) models with $a_* \gtrsim 0.98$ that reproduce the
millimeter flux have a hot, dense inner disk that would overproduce
X-rays via inverse Compton scattering.  Lower spin, $a_* \sim 0.9$,
models are therefore favored; (3) the observed source size and flux fix
the temperature of the emitting electrons $T_e = F_\nu c^2/(4\pi k \nu^2
\sigma^2)$ ($\sigma \equiv$ the RMS size of the source on the sky) and
this turns out to favor $T_i/T_e \simeq 3$ models.  The ``best-bet''
model from \citet{moscibrodzka:2009} has $a_* \simeq 0.94$, $i = 85$deg,
$\dot{M} \equiv \dot{M}_0 \simeq 2 \times 10^{-9} {\rm M_{\odot}
yr^{-1}}$, and $T_i/T_e = 3$.

The relativistic disk model uses self-consistent dynamics and radiative
transfer but is not unique. The electron distribution function is
particularly poorly constrained.  It is likely anisotropic, may contain
multiple temperature components (\citealt{rique:2012}) and
power-law components, and vary in basic functional form with time and
position.  Alternative accretion models (e.g. \citealt{broderick:2011},
\citealt{roman:2010a}) make different assumptions about the flow and/or
distribution function and favor slightly different $\dot{M}, a_*,$ and
$i$.  These models may respond differently to an increase in $\dot{M}$.

Indeed, radically different models may also fit the data.  The model
dynamics depends on the initial magnetic field distribution,
particularly the distribution of vertical flux through the accretion
disk; models with large vertical magnetic flux (e.g.
\citealt{mckinney:2012}) are likely to respond quite differently to
variations in the mass flux. Also, jet models for Sgr~A*
(\citealt{falcke:2000jet}, \citealt{loeb:2007}, \citealt{falcke:2009})
posit a luminous jet and a comparatively dim accretion disk. Again,
these may respond differently to an influx of mass.

How, then, does our relativistic disk model respond to an increase in
$\dot{M}$?  In this {\em Letter} we calculate the $1.3$ and $0.87$mm
flux and source size as well as the spectrum that would result for the
best-bet model over a range of $\dot{M}$. One key question we
seek to answer is whether a small increase in $\dot{M}$ would
hide the event horizon (and the signature ring-like appearance of the
photon orbit, also known as the shadow or silhouette of the event
horizon) underneath a synchrotron photosphere. This might prevent
detection of the photon orbit by the planned Event Horizon Telescope
\citep{doeleman:2008}. Another key question is whether the increased
$\dot{M}$ would make Sgr~A* detectable in its quiescent state in
the IR and X-ray. Below we describe variation of the flux and source
morphology at $230$GHz ($1.3$mm) and $345$GHz ($0.87$mm)
(\S~\ref{results}), describe variation of the spectrum
(\S~\ref{spectra}), and finally discuss which features of the
results are likely to be most robust (\S~\ref{discussion}).

\section{Change of Sgr~A* sub-mm luminosity and size for enhanced $\dot{M}$}\label{results}

How do we naively expect the mm disk image size and flux to respond to
changes in $\dot{M}$?  In our model, $1.3$mm emission in Sgr~A* is
thermal synchrotron emission from plasma with optical depth $\tau_\nu
\sim 1$, near the ISCO. The true electron distribution undoubtedly
contains nonthermal components (see, Riquelme et al. 2012).
Models with thermal + power-law distribution functions (e.g.
\citealt{broderick:2011}) contain an O(1/3) nonthermal contribution to
the flux at $1.3$mm, which hints at how uncertainty in the distribution
function translates into uncertainties in the spectrum.

The thermal synchrotron absorptivity is $\alpha_{\nu,a}=j_\nu/B_\nu$,
where $j_{\nu} = \frac{\sqrt{2}\pi e^2 n_e \nu_s}{3cK_2(\Theta_e^{-1})}
(X^{1/2} + 2^{11/12} X^{1/6} )^2 \exp(-X^{1/3})$, $X = \nu/\nu_s$,
$\nu_s=2/9 (eB/2\pi m_ec) \Theta_e^2 \sin\theta$, $\theta$ is an angle
between the magnetic field vector and emitted photon, $K_2$ is a
modified Bessel function of the second kind \citep{leung:2011} and
$B_\nu \simeq 2 \nu^2 \Theta_e m_e$.  Near $1.3$mm, $X \sim 1$, and the
emissivity is nearly independent of frequency, so $j_\nu \propto \nu^0
n_e B$ and $\alpha_{\nu,a} \propto \nu^{-2} n_e B \Theta_e^{-1}$.

We will assume that $n_e \propto \dot{M} r^{-3/2}$, $\Theta_e \propto
1/r$, and $\beta \sim$ const., so that $B \propto \dot{M}^{1/2}
r^{-5/4}$ for $r > G M/c^2$, and ignore relativistic corrections.  Then
for $\tau_\nu \ll 1$ (or $\dot{M} \ll \dot{M}_0$) the source has size
$\sim G M/c^2$ and the flux $F_\nu \sim (4/3)\pi (G M/c^2)^3 j_\nu
\propto \dot{M}^{3/2}$.  For $\tau_\nu \gg 1$ the source size is set by
the photosphere radius $r_{ph}$ where $\int_{r_{ph}}^\infty \alpha_a(r)
dr = 1$ (i.e. for $\dot{M} \gg \dot{M}_0$, but not so large that
$\Theta_e(r_{ph}) < 0.5$ so that our emissivity approximation fails).
Then $r_{ph} \propto \dot{M}^{3/2}/\nu^2$, and the flux $F_\nu \propto
r_{ph}^2 B_\nu(r_{ph}) \sim \dot{M}^{9/4}/\nu$.

These simple scaling laws, unfortunately, are not a good description of
the variation of source size and flux with $\dot{M}$.  There are at
least three reasons for this.  First, relativistic effects are very
important; for $\dot{M} \sim \dot{M}_0$ emission comes from close to the
photon orbit and the source size is determined by Doppler beaming and
gravitational lensing. Second, for the best-bet model at $1.3$mm
$\tau_\nu \sim 1$, so in a turbulent flow there is a complicated
variation of the size of the effective photosphere with $\dot{M}$.
Third, the emissivity is not precisely frequency independent near peak.
We therefore need to turn to numerical models.

The best-best model is taken from a survey of 2D models.  Here we adopt
the best-bet model parameters ($a_*=0.94$, $i=85 \deg$ and $T_i/T_e=3$)
and use them to set parameters for a fully 3D model
\citep{shiokawa:2011}~\footnote{A 3D GRMHD model parameter survey is
still too computationally expensive.  It also has a poor return on
investment given the electron distribution function uncertainties.}.  We
use the same data set as \citet{dolence:2012}, and choose three
representative snapshots taken at times when the flow is quiescent
($t=5000, 9000$ and $13000 GM/c^3$, where $GM/c^3 = 20$s).  We then
recalculate disk images and spectra for $\dot{M} = (0.5, 1, 2, 4, 8, 16,
32,64) \dot{M}_0$.

The 3D model with $\dot{M} = \dot{M}_0$ is broadly consistent with
observational data but is slightly more luminous at higher energies than
2D models ($\beta$ is lower in the 3D models, and this changes the X-ray
to millimeter color).  The model is self-consistent only for $\dot{M} <
64 \dot{M}_0$.  At higher $\dot{M}$ the efficiency of the flow is $>
0.1$ and therefore our neglect of cooling in the underlying GRMHD model
is not justified.  At higher $\dot{M}$ the $1.3$mm photosphere also lies
outside the limited range in radius where $\langle d\dot{M}/dr \rangle =
0$, so the flow is not in a steady state.

The images and total fluxes emitted by the disk at $230$ and $345$GHz
are calculated using a ray tracing scheme \citep{noble:2007}.  To
estimate the size of the emitting region we calculate the eigenvalues of
the matrix formed by taking the second angular moments of the image on
the sky (the principal axis lengths).  The major and minor axis
eigenvalues, $\sigma_1$ and $\sigma_2$ respectively, are related by
$\sigma={\rm FWHM}/2.3$ to the FWHM of the axisymmetric Gaussian model
used to interpret the VLBI observations.  We use $\langle \sigma \rangle
= (\sigma_1+\sigma_2)/2$ to measure the average radius of the emitting
spot.

\begin{figure*}[ht!]
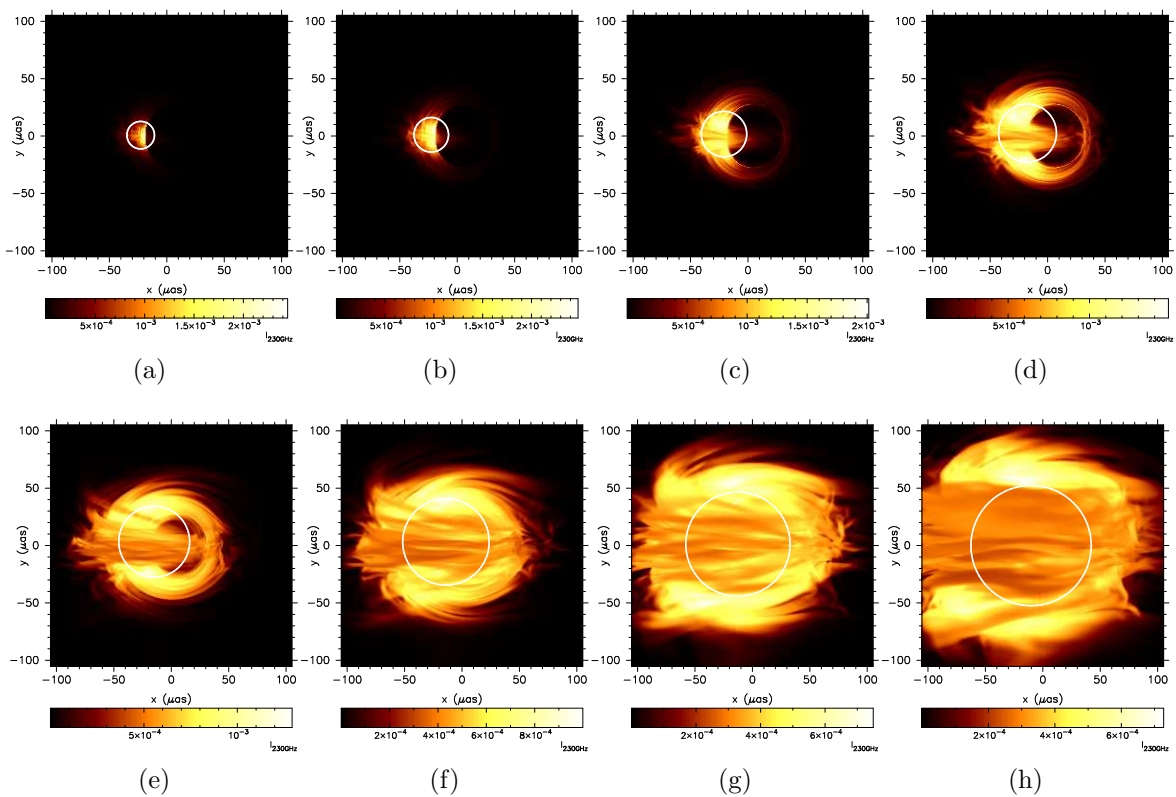

\centering
\subfloat[]{\label{fig:1a}\includegraphics[width=0.27\textwidth,angle=-90]{fig1a.ps}}
\subfloat[]{\label{fig:1b}\includegraphics[width=0.27\textwidth,angle=-90]{fig1b.ps}}
\subfloat[]{\label{fig:1c}\includegraphics[width=0.27\textwidth,angle=-90]{fig1c.ps}}
\subfloat[]{\label{fig:1d}\includegraphics[width=0.27\textwidth,angle=-90]{fig1d.ps}}\\
\subfloat[]{\label{fig:1e}\includegraphics[width=0.27\textwidth,angle=-90]{fig1e.ps}}
\subfloat[]{\label{fig:1f}\includegraphics[width=0.27\textwidth,angle=-90]{fig1f.ps}}
\subfloat[]{\label{fig:1g}\includegraphics[width=0.27\textwidth,angle=-90]{fig1g.ps}}
\subfloat[]{\label{fig:1h}\includegraphics[width=0.27\textwidth,angle=-90]{fig1h.ps}}
\caption{Horizon silhouette detectability at 230 GHz for various
$\dot{M}$. Panels from a) to h) show the images
  of Sgr~A* calculated for $\dot{M} =(0.5, 1, 2, 4, 8, 16, 32, 64) \dot{M}_0 $,
  respectively. The center of the circle is positioned at the image centroid
  and its radius, $r=(\sigma_1+\sigma_2)/2$ is the RMS radius of
  the emitting region.\label{fig:1}}
\end{figure*}

\begin{figure*}[ht!]
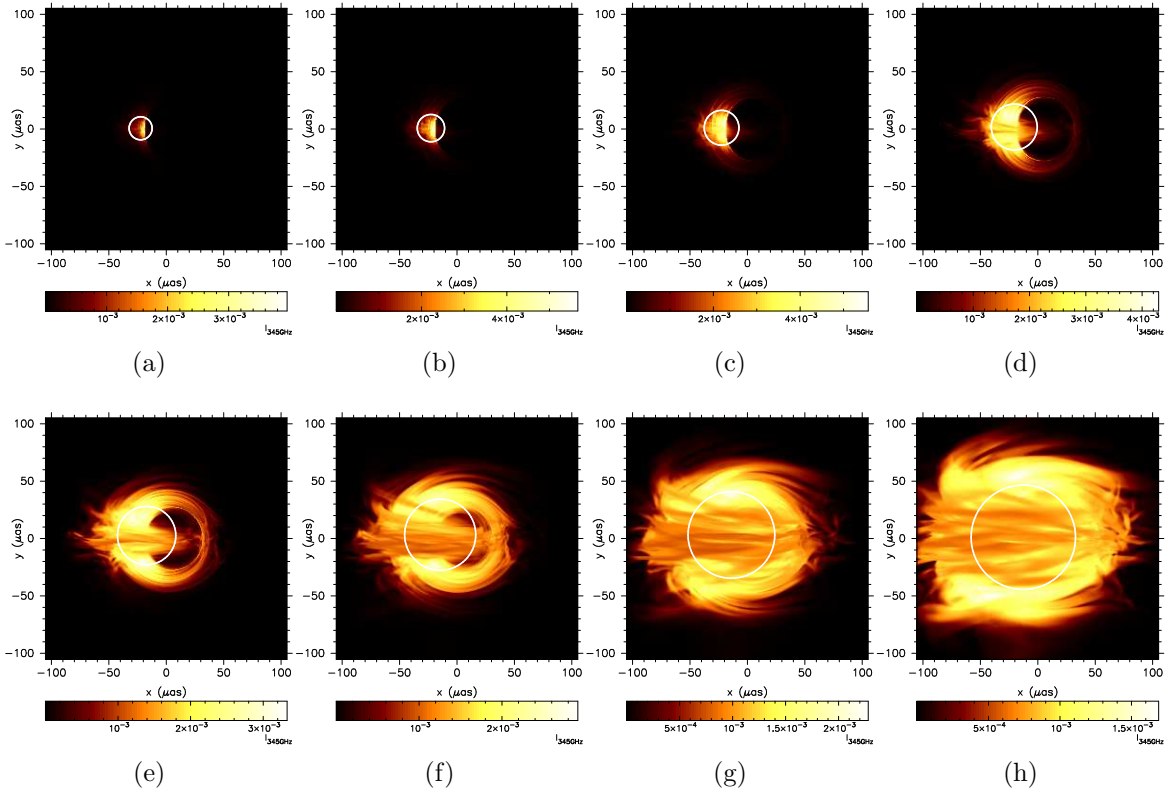

\centering
\subfloat[]{\label{fig:2a}\includegraphics[width=0.27\textwidth,angle=-90]{fig2a.ps}}
\subfloat[]{\label{fig:2b}\includegraphics[width=0.27\textwidth,angle=-90]{fig2b.ps}}
\subfloat[]{\label{fig:2c}\includegraphics[width=0.27\textwidth,angle=-90]{fig2c.ps}}
\subfloat[]{\label{fig:2d}\includegraphics[width=0.27\textwidth,angle=-90]{fig2d.ps}}\\
\subfloat[]{\label{fig:2e}\includegraphics[width=0.27\textwidth,angle=-90]{fig2e.ps}}
\subfloat[]{\label{fig:2f}\includegraphics[width=0.27\textwidth,angle=-90]{fig2f.ps}}
\subfloat[]{\label{fig:2g}\includegraphics[width=0.27\textwidth,angle=-90]{fig2g.ps}}
\subfloat[]{\label{fig:2h}\includegraphics[width=0.27\textwidth,angle=-90]{fig2h.ps}}
\caption{Same as in Figure~\ref{fig:1} but for $\nu=345$ GHz.\label{fig:2}}
\end{figure*}

Figures~\ref{fig:1} and~\ref{fig:2} show the variation of $230$ and
$345$ GHz Sgr~A* model images with $\dot{M}$ (based on a single snapshot
from the 3D GRMHD simulation).  Evidently modest increases $\dot{M}$
will make the ring-like feature that is the observational signature of
the photon orbit easier to detect.  For $\dot{M} \gtrsim 8 \dot{M}_0$,
however, the ring (or black hole silhouette) is hidden beneath the
synchrotron photosphere at $230$ GHz.  The silhouette survives for
higher $\dot{M}$ at $345$ GHz, disappearing only at $\dot{M} \gtrsim 16
\dot{M}_0$.  For low $\dot{M}$ the silhouette is also difficult to
detect because the emitting region is too small. 

\begin{figure*}[ht]
\centering
\subfloat[]{\label{fig:3a}\includegraphics[width=0.5\textwidth]{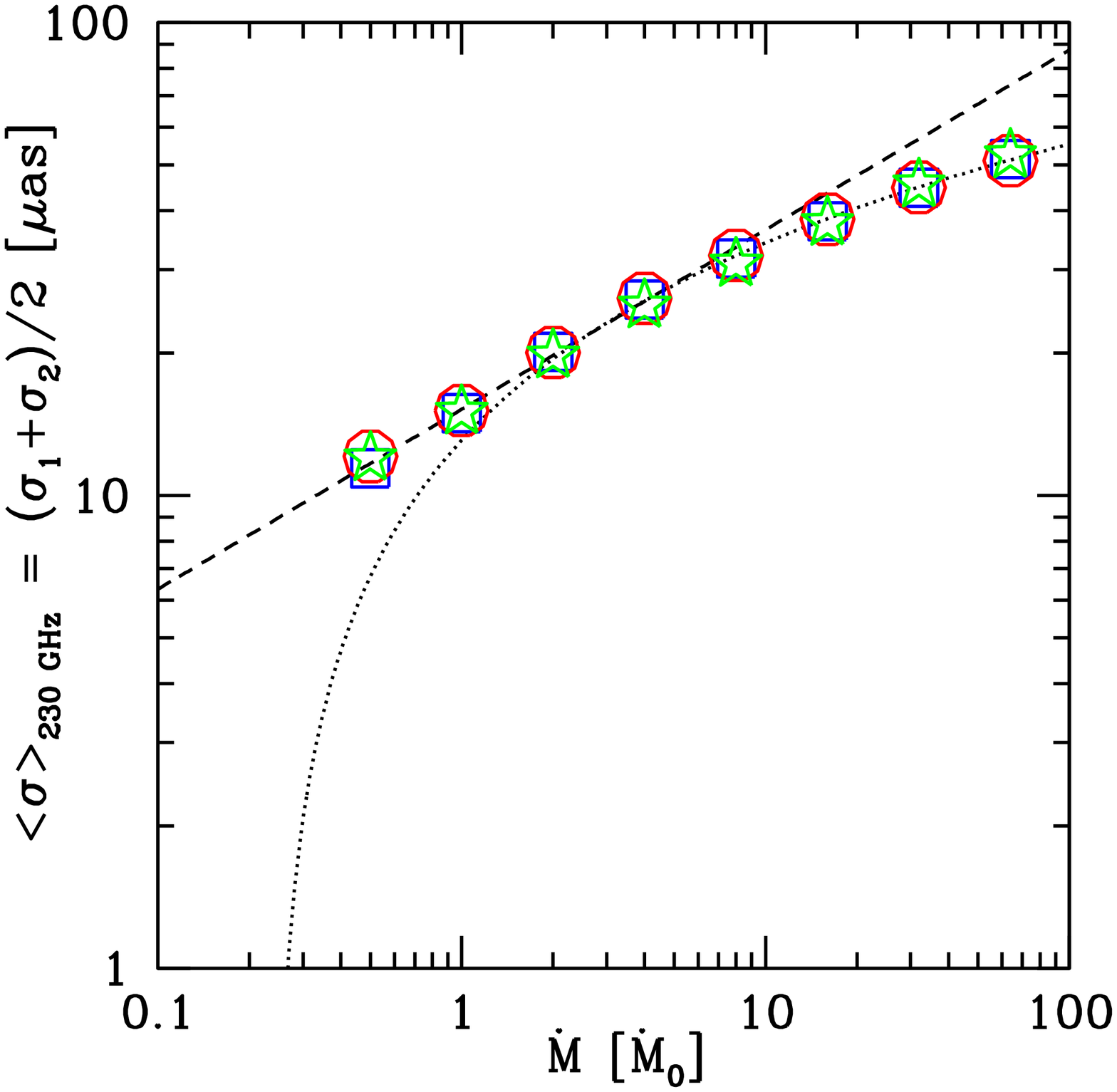}}
\subfloat[]{\label{fig:3b}\includegraphics[width=0.5\textwidth]{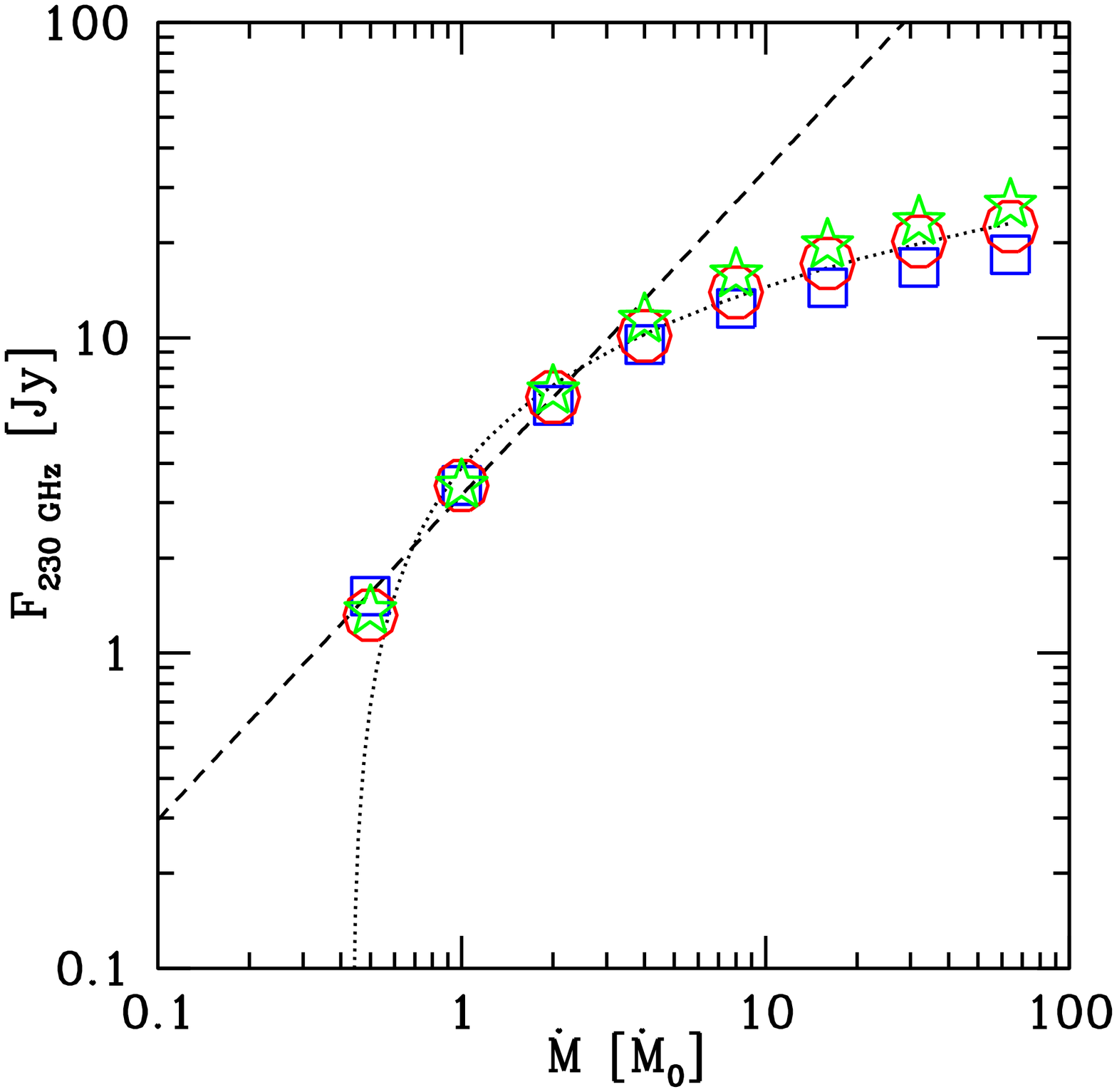}}\\
\subfloat[]{\label{fig:3c}\includegraphics[width=0.5\textwidth]{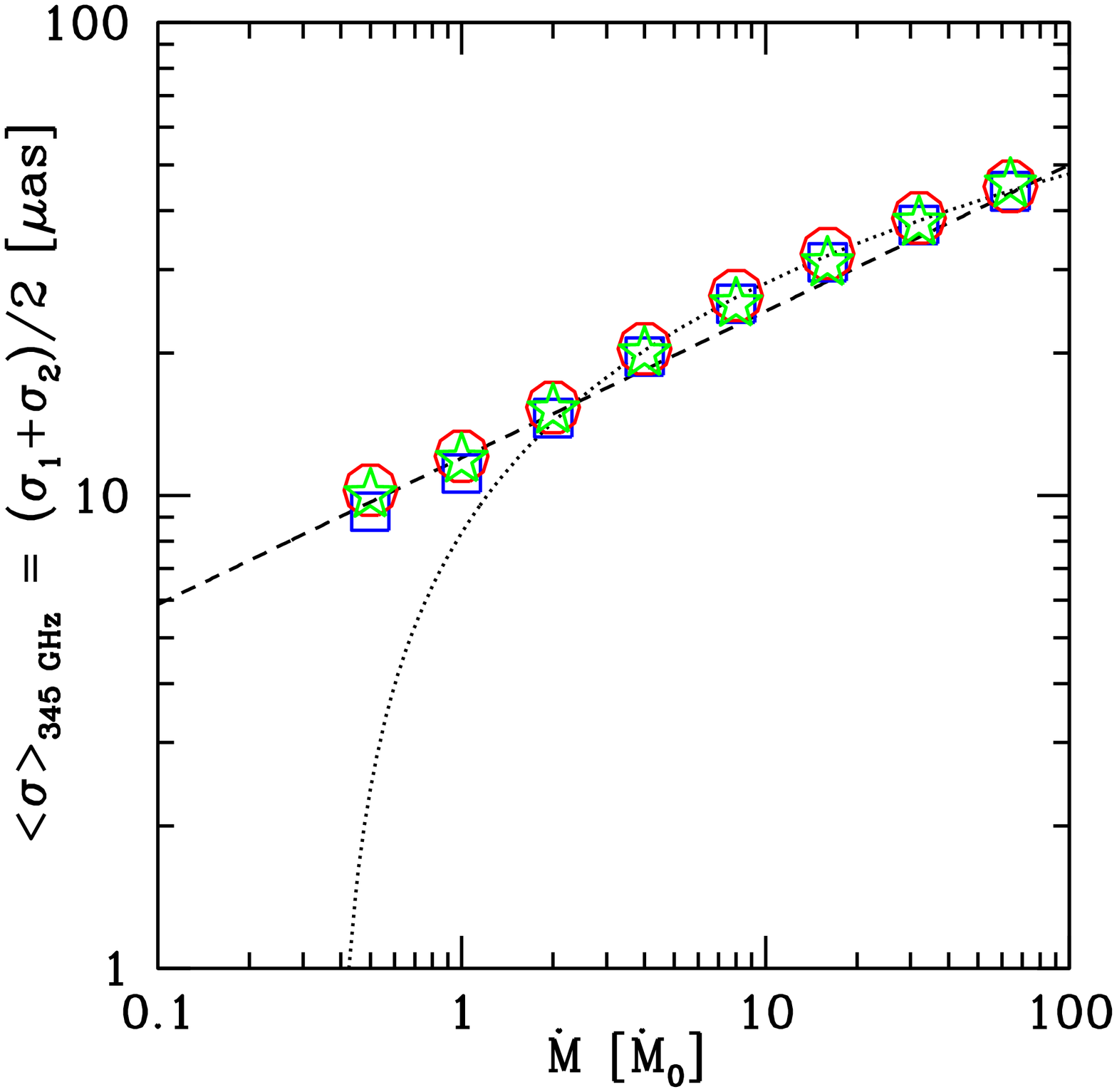}}
\subfloat[]{\label{fig:3d}\includegraphics[width=0.5\textwidth]{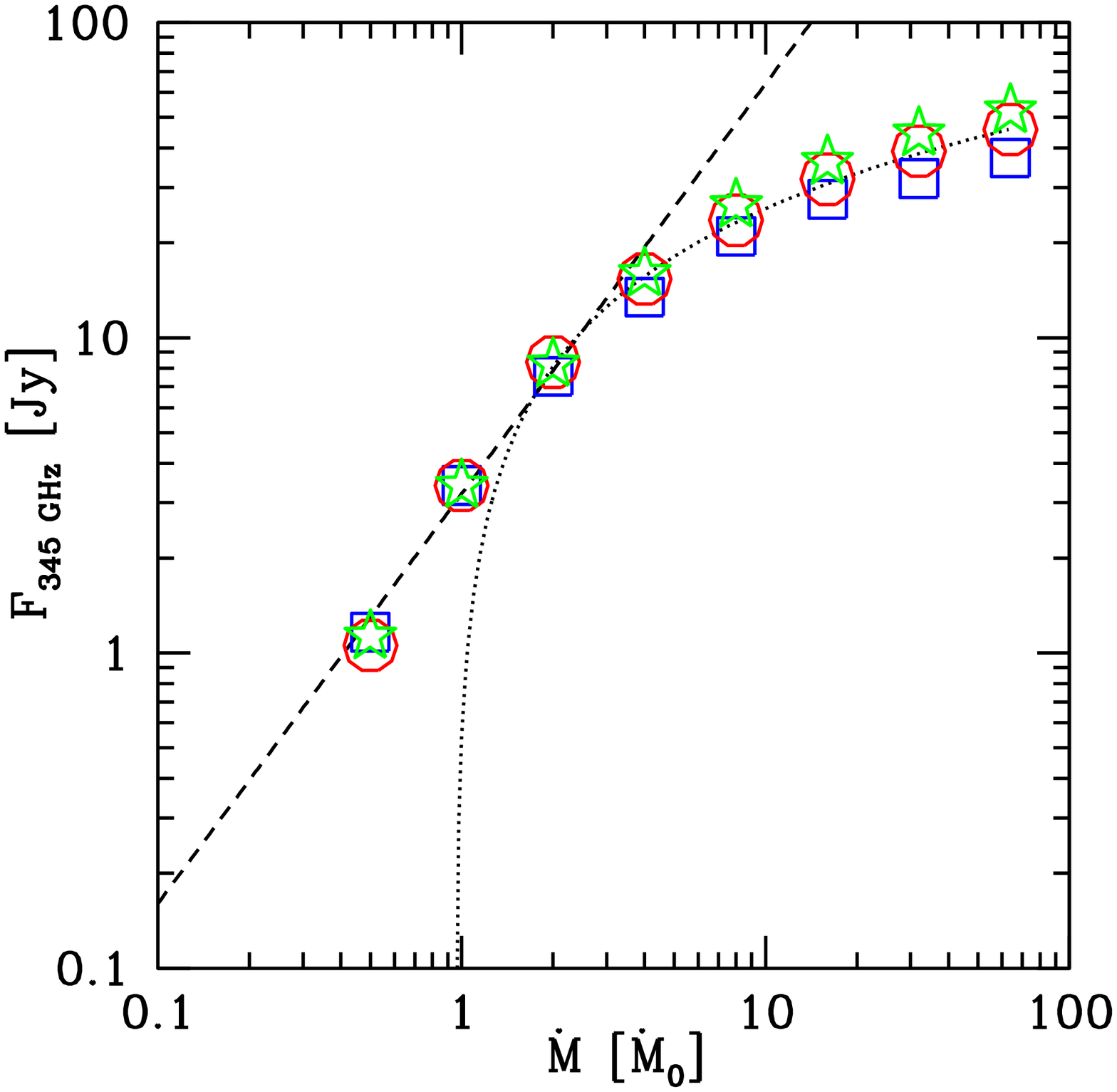}}
\caption{
Sizes (in terms of $\sigma$) of the emitting region and the total flux as a
function of $\dot{M}$ at $230$ (upper panels) and $345$GHz (lower
panels). Different point types correspond to three different snapshots from
the 3D GRMHD simulation. The dashed/dotted lines show the best fit to the data
by Equations~\ref{fitsig230},~\ref{fitsig345}, \ref{fitF230}, and
\ref{fitF345}.
\label{fig:3}}
\end{figure*}

Figure~\ref{fig:3}, shows the accretion flow image size
($\langle\sigma\rangle$, circle radii in Figure~\ref{fig:1}
and~\ref{fig:2}) and flux at $230$ and $345$GHz as a function of
$\dot{M}$. Different point types correspond to three different snapshots
from the 3D GRMHD simulation.  The size of the emission region and flux
increase with $\dot{M}$.  The 345/230 GHz flux ratio increases with the
increasing $\dot{M}$; this is caused by the shift of the synchrotron
peak towards higher energies at higher $\dot{M}$.  

The following simple fitting formulas describe how $\sigma$ and $F_\nu$
depend on $\dot{M}$:
\begin{equation}
\langle \sigma \rangle_{230{\rm GHz}}=
\begin{cases} 
15.2 \times(\frac{\dot{M}}{\dot{M}_0})^{0.38},& \mbox{for}
\frac{\dot{M}}{\dot{M}_0} < 2\\
21.1 \times \log_{10} (\frac{\dot{M}}{\dot{M}_0})+ 13,& \mbox{for} \frac{\dot{M}}{\dot{M}_0} \geq 2
\end{cases}
\, \, {\rm [\mu as]}
\label{fitsig230}
\end{equation}

\begin{equation}
\langle \sigma \rangle_{345{\rm GHz}}=
\begin{cases} 
12 \times (\frac{\dot{M}}{\dot{M}_0})^{0.31},& \mbox{for}
\frac{\dot{M}}{\dot{M}_0} < 2\\
19.7 \times \log_{10} (\frac{\dot{M}}{\dot{M}_0})+ 8.3,& \mbox{for} \frac{\dot{M}}{\dot{M}_0} \geq 2
\end{cases}
\, \, {\rm [\mu as]}
\label{fitsig345}
\end{equation}

\begin{equation}
F_{230{\rm GHz}}=
\begin{cases} 
3.17 \times (\frac{\dot{M}}{\dot{M}_0})^{1.03},& \mbox{for}
\frac{\dot{M}}{\dot{M}_0} < 2\\
10.62 \times \log_{10} (\frac{\dot{M}}{\dot{M}_0})+ 3.8,& \mbox{for} \frac{\dot{M}}{\dot{M}_0} \geq 2
\end{cases}
 \, \, {\rm [Jy]} \label{fitF230}
\end{equation}

\begin{equation}
F_{345{\rm GHz}}=
\begin{cases} 
3.2 \times (\frac{\dot{M}}{\dot{M}_0})^{1.3},& \mbox{for}
\frac{\dot{M}}{\dot{M}_0} < 2\\
25.13 \times \log_{10} (\frac{\dot{M}}{\dot{M}_0})+ 0.54,& \mbox{for} \frac{\dot{M}}{\dot{M}_0} \geq 2
\end{cases}
 \, \, {\rm [Jy]} \label{fitF345}
\end{equation}
The above constants are nontrivial to interpret because they encapsulate
the complexities of the accretion flow structure and relativistic
effects in the radiation transport. The advantage of the above formulas
is their simplicity.  The fitting functions are shown in
Figure~\ref{fig:3} as dashed and dotted lines.

\begin{figure}[ht]
\epsscale{1.0}
\plotone{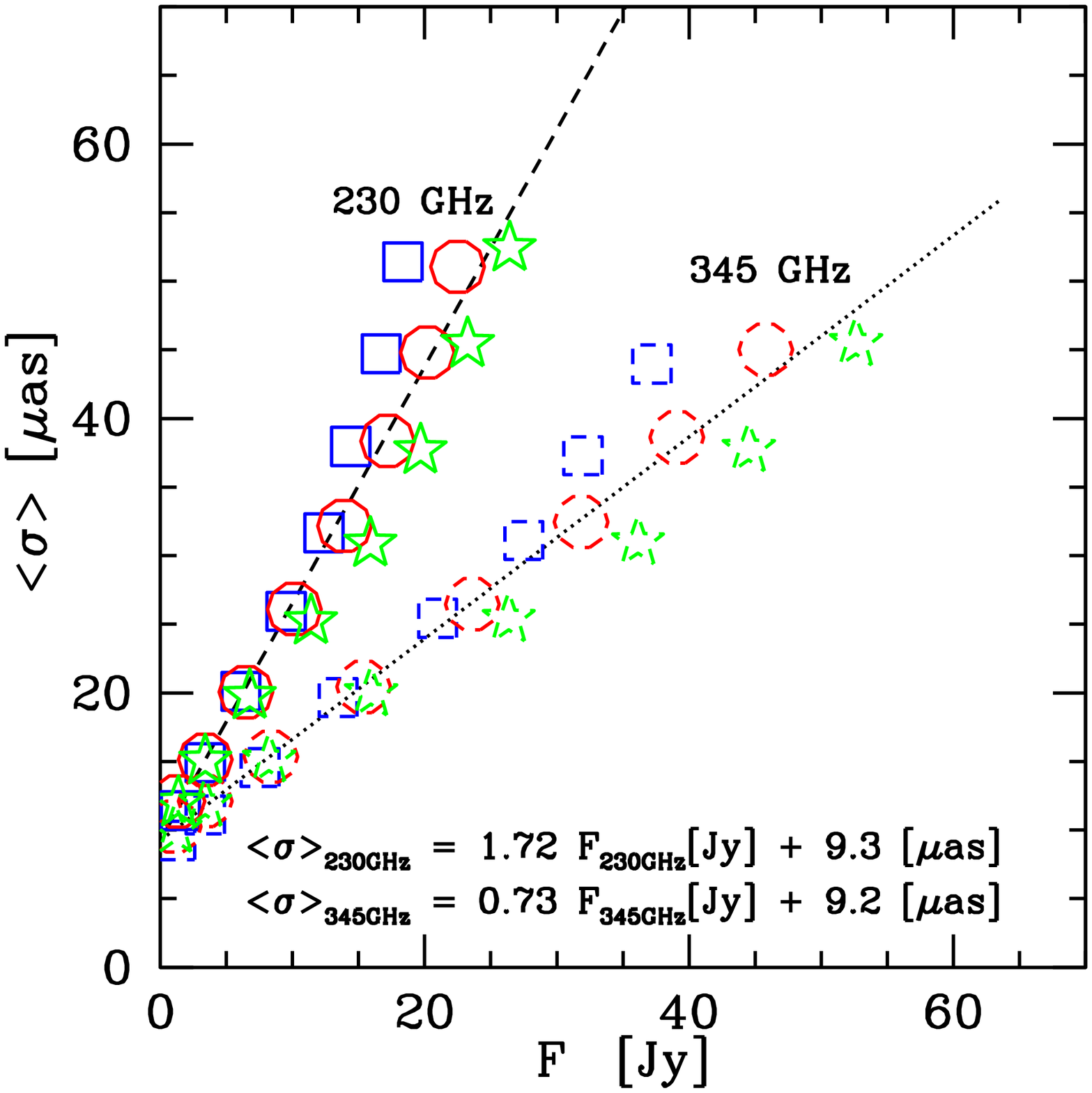}
\caption{Relation between two observables: the flux and the size of the 
image at $\nu=$230 and 345 GHz. The dashed and dotted lines are the 
best fits to the data points.
\label{fig:4}}
\end{figure}
Figure~\ref{fig:4} shows the relation between two observables, $\sigma$
and $F_\nu$. The size is a linear function of the flux and increases
more steeply at 230 GHz than at 345 GHz. We also provide two
phenomenological scaling laws fitted to the data:
\begin{equation}
\langle \sigma \rangle_{230GHz} = 1.72 (F_{230{\rm GHz}}/{\rm Jy}) + 9.3 \,\,\,{\rm [\mu as]}  \label{form1}
\end{equation}
\begin{equation}
\langle\sigma\rangle_{345GHz} = 0.73 (F_{345{\rm GHz}}/{\rm Jy}) + 9.2 \,\,\,{\rm [\mu as]}  \label{form2}
\end{equation}
Notice that these fits apply to the best-bet model only.  For other
$a_*$, $T_i/T_e$, or $i$ the scalings will be slightly different,
although if the source is optically thick the linear scaling follows
directly from $T_e \propto 1/r$.

Notice that Equation~\ref{form1} predicts the change of Sgr~A* size
by 9 per cent when 230 GHz flux changes from 2 to 2.7 Jy.  Taking into
account observational and theoretical uncertainties this is consistent
with the observed variations of the size of the source which increases
by a few per cent as the flux increases from 2 to 2.7 Jy
\citep{fish:2011}.

\section{Spectra}\label{spectra} 

Our model spectra are generated by thermal synchrotron emission in the
submillimeter/far-IR bump, and by Compton scattering in the X-rays.  The
spectral slope of flaring NIR emission, and its high degree of linear
polarization \citep{dodds-eden:2011}, imply that it is synchrotron from
a small, nonthermal tail of high energy electrons that is not (but can
be; see \citealt{leung:2010}) included in our best-bet model.

What are the expected scaling laws?  Again, $j_\nu \sim n_e B \sim
\dot{M}^{3/2}$. The luminosity around the synchrotron peak is $L_{peak}
\sim 4 \pi \nu_{peak} j_{\nu_{peak}} (GM/c^2)^3 \sim \dot{M}^{9/4}$,
where $\nu_{peak} \sim \dot{M}^{3/4}$ is such that $\alpha_\nu GM/c^2 =
1$. 

The emission rightward of the MIR/NIR is produced by Compton
up-scattered synchrotron radiation. The Thomson depth $\tau_{sc} =
\sigma_{TH} n_e GM/c^2 \sim \dot{M}$, the X-ray luminosity is expected
to scale as: $\nu L_\nu (\nu \approx 10^{18} {\rm Hz} ) \sim L_{peak}
\tau_{sc} \sim \dot{M}^{13/4}$, assuming that X-rays are produced
primarily by singly scattered synchrotron photons.

\begin{figure}[ht]
\epsscale{1.0}
\plotone{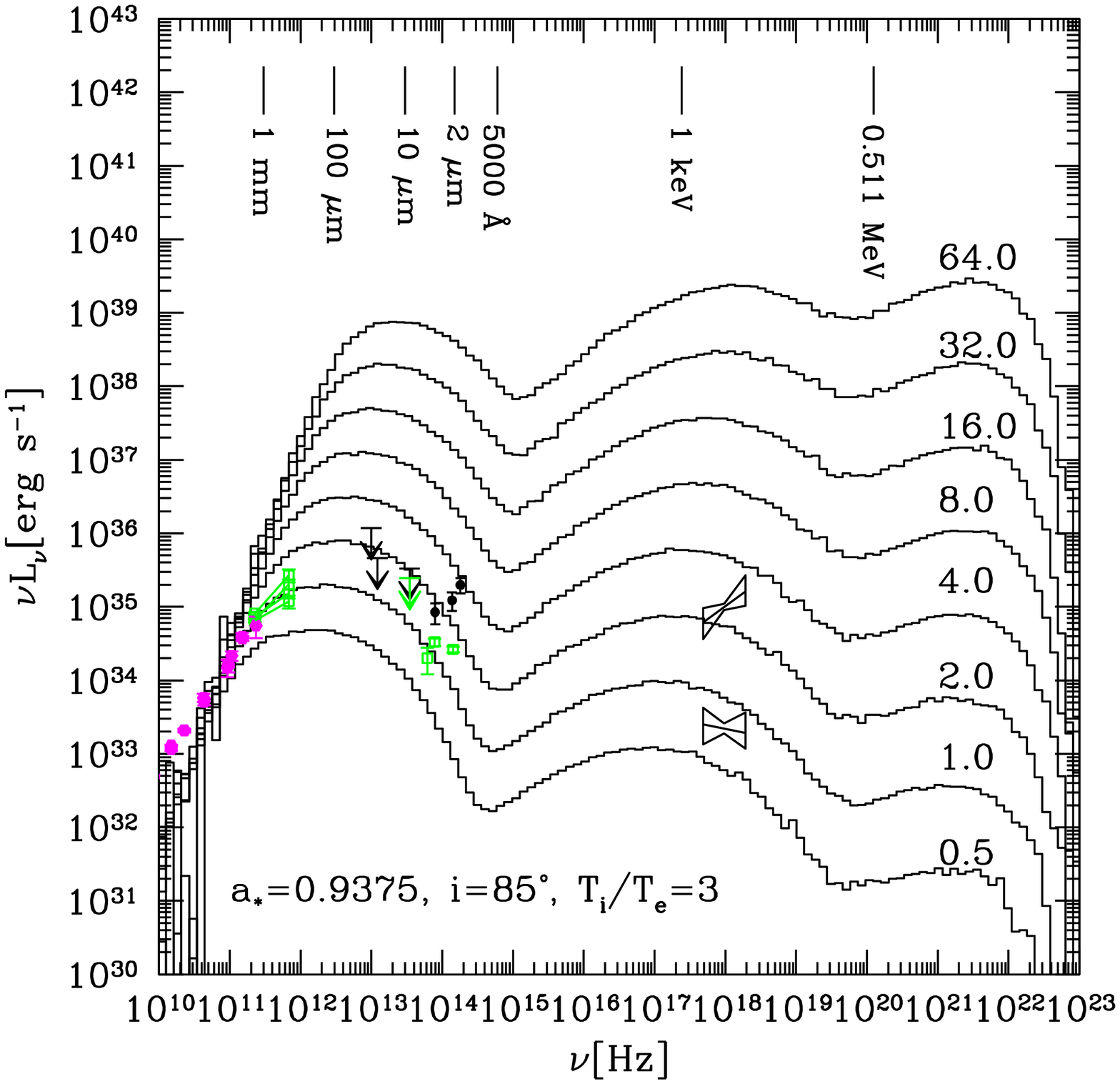}
\caption{Spectrum emitted by 3D disk model for various $\dot{M}$.  The
$\dot{M}/\dot{M}_0$ is shown on the righthand side.  Observational points
and upper limits are taken from: \citealt{falcke:1998}, \citealt{an:2005},
\citealt{marrone:2006a}, \citealt{melia:2001}, \citealt{schoedel:2011},
\citealt{baganoff:2003}. The black symbols in the NIR showing the flaring
state are from \citealt{genzel:2003}. An example of X-ray flare is taken from \citealt{baganoff:2001}.
\label{fig:5}}
\end{figure}

What do the numerical models show?  Figure~\ref{fig:5} shows spectra
emitted from the 3D disk model as observed at $i\approx 85 \deg$. The
SEDs are calculated using a general relativistic Monte Carlo scheme
\citep{dolence:2009}. The NIR luminosity ${\nu}L_{\nu}(\nu=10^{14} {\rm
Hz}) \sim \dot{M}^{2.5}$, which is only slightly steeper than the
expected dependence for the synchrotron peak.
${\nu}L_{\nu}(\nu\approx10^{18} {\rm Hz}) \sim \dot{M}^{3.25}$ agrees
well with the expected scaling.

We conclude that Sgr~A* would become a persistent MIR and X-ray source
(above the present upper limits of $84 {\rm mJy}$ in MIR and $2.4 \times
10^{33} {\rm erg s^{-1}}$ in X-rays) if $\dot{M} > 2 \dot{M}_0$.  This
is conservative in the sense that our models are strictly thermal. The
addition of a high energy nonthermal tail would only increase the
MIR/NIR and X-ray flux.

We do not consider higher accretion rate models because for $\dot{M} =
64 \dot{M}_0$ (${\nu}L_{\nu}(\nu=10^{18} {\rm Hz})=10^{39} {\rm erg
s^{-1}}$) the model becomes radiatively efficient,
$\epsilon=L_{Bol}/\dot{M} c^2 > 0.1$, and our neglect of radiative
cooling in the underlying 3D GRMHD model is unjustified.

Finally, notice that the MeV flux increases sharply with $\dot{M}$.
This suggests that electron-positron pair production by photon-photon
collisions in the funnel over the poles of the hole would increase
sharply (the pair production $\dot{n}_{e^\pm} \sim L_{\gamma}^2$,
\citealt{moscibrodzka:2011}). If this pair production is connected to
jet production, it is reasonable to think that a high $\dot{M}$ Sgr~A*
might also produce a jet.

\section{Discussion}\label{discussion}

In summary, we have used general relativistic disk models and
relativistic radiative transfer to recalculate millimeter images and
spectra for Sgr~A* at a range of $\dot{M} > \dot{M}_0$.  Our models
predict the following: (1) if the 230 GHz flux increases by more than a
factor of 2, corresponding to an increase of $\dot{M}$ by more than a
factor of 2, the central accretion flow will become a persistent,
detectable, MIR and X-ray source; (2) the photon orbit, which produces
the narrow ring of emission visible in Figures 1 and 2, becomes easier
to detect for modest increases in the 230 GHz flux; (3) the photon orbit
is cloaked beneath the synchrotron photosphere at $230$ GHz for $\dot{M}
\gtrsim 8 \dot{M}_0$, or 230 GHz flux $\gtrsim 13$ Jy; (4) the photon
orbit is cloaked at $345$ GHz only at higher $\dot{M} \gtrsim 16
\dot{M}_0$, or 230 GHz flux $\gtrsim 17$ Jy; (5) the size of the source
increases in proportion to the flux at both $230$ and $345$ GHz.  We
suspect that almost any accretion model for Sgr~A* with a spatially
uniform model for the plasma distribution function will reach
qualitatively similar conclusions, but that jet models may differ
significantly.  There are order-unity uncertainties in our model
predictions due to uncertainties in the plasma model.

What range of changes in $\dot{M}$ are reasonable?  In our best-bet
model the mass at radii within a factor of two of the pericenter radius
is $\approx 10^{-2.5}$ $M_{\oplus}$, assuming steady mass inflow from
$r_p$ to the event horizon. The addition of even a fraction of the
inferred cloud mass to the accretion flow in a ring near $r_p$ would
(eventually) increase $\dot{M}$ by a factor of $\sim 300$.  On the other
hand, stellar winds supply mass in the neighborhood of the central black
hole at $\sim 10^{-3} M_{\odot} yr^{-1}$. Models by
\citet{quataert:2004} and \citet{roman:2010b} suggest that most of this
mass is ejected in the form of a wind, and that $\sim 10^{-4.5}
M_{\odot} yr^{-1}$ to $10^{-7.3} M_{\odot} yr^{-1}$ flows inward.  A
reasonable extrapolation of these models suggest the accretion flow at
$r < r_p$ has a mass of $\sim 2 M_{\oplus}$; this is comparable to
estimates of the mass of the inflowing cloud, so in this case we might
expect a factor of $2$ increase in $\dot{M}$.

\acknowledgments

This work was supported by the National Science Foundation under grant
AST 07-09246 and by NASA under grant NNX10AD03G, through TeraGrid
resources provided by NCSA and TACC.

%\bibliographystyle{apj}
%\bibliography{local}

\end{document}